\documentclass{cernyrep}
\usepackage{texnames}
\usepackage[T1]{fontenc}
\usepackage[colorlinks=true, pdftex]{hyperref}

\newcommand{\bi}{\begin{itemize}}
\newcommand{\ei}{\end{itemize}}
\newcommand{\bc}{\begin{center}}
\newcommand{\ec}{\end{center}}
\newcommand{\be}{\begin{equation}}
\newcommand{\ee}{\end{equation}}
\newcommand{\beqna}{\begin{eqnarray}}
\newcommand{\eeqna}{\end{eqnarray}}
\newcommand{\nn}{\nonumber}

\begin{document}
  \title{Cosmological inflation}
\author{K. Enqvist}
\institute{Physics Department, University of Helsinki, and Helsinki Institute of Physics, Finland}
\maketitle 

\begin{abstract}
The very basics of cosmological inflation are discussed. We derive the
equations of motion for the inflaton field, introduce the slow-roll
parameters, and
present the computation of the inflationary perturbations
and their connection to the temperature fluctuations of
the cosmic microwave background.
\end{abstract}

\section{Introduction}

Cosmological inflation is, by definition, a period of superluminal expansion in the very early
universe\cite{lythliddle}. In practice, the rate of expansion is usually taken to be (quasi)exponential. Superluminal expansion
does not contradict the theory of relativity, which states that no
\emph{signal} can propagate faster than the speed of light. One cannot use the expansion of space to
send any signal, and as we will see, superluminal expansion is indeed a solution to the field equations
of general relativity.

Historically, inflation was introduced to solve certain fine-tuning issues in the hot Big Bang scenario.
However, its main attraction turned out to provide a mechanism for the origin of the density perturbations
required for structure formation. Density perturbations leave their imprint on the cosmic microwave
background (CMB) and were first detected by the COBE satellite \cite{cobe}.

The simplest starting point for cosmology is to assume that the universe is homogeneous and isotropic.
It can be shown that the most general metric describing  such a Friedmann--Robertson--Walker (FRW) universe
reads
\be\label{frwmetric}
ds^2=dt^2-a^2(t)\left(\frac{dr^2}{1-kr^2}+r^2d\Omega^2\right)~,~~k=\pm1,0~,
\ee
where $k$ is the spatial curvature parameter.
Here the convention is such that today the scale factor $a(t_{\rm now})=1$. Within the FRW framework, matter is homogeneous
and continuous. Therefore, it can be described by the energy-momentum tensor of a perfect fluid, given by
\be\label{tmunu}
T_{\mu\nu}=(\rho+p) u_\mu
u_\nu-pg_{\mu\nu}~.
\ee
Here $\rho$ is the energy density and $p$ the pressure of the fluid.
In the rest frame of the fluid, where the four-velocity is given by $u=(1,0)$,
we find $T_{00}=\rho,~T_{ii}=p$.
One then substitutes the
metric (\ref{frwmetric}) and the energy-momentum tensor (\ref{tmunu}) into the Einstein equations,
which yield evolution equations for $a(t)$ and the components of the energy-momentum tensor:
\begin{eqnarray}
\label{friedmanni}
H^2\equiv\left(\frac{\dot a}{a}\right)^2&=&\frac{8\pi G}{3}\rho-\frac{k}{a^2}~,\\
\label{conteq}
\dot\rho+3H(p+\rho)&=&0~,
\end{eqnarray}
where $G$ is the gravitational constant
and $H$ is the Hubble parameter. These must be supplemented by an equation of state $p=w\rho$, whence it
follows from (\ref{conteq}) that $\rho\propto a^{-3(w+1)}$. For cold dust, $w=0$; for relativistic particles
(radiation), $w=1/3$.

If $k=0$, the geometry is said to be flat. By virtue of (\ref{friedmanni}), in such a case the energy density is directly related to the Hubble rate:
\be
\rho=\frac{3H^2}{8\pi G}\equiv\rho_c~,
\ee
where $\rho_c$ is called the critical density. Usually the densities are written in units of the critical density and are then denoted
by
\be
\Omega=\rho/\rho_c~.
\ee
From the Friedmann equation (\ref{friedmanni}) we then find that the difference from the critical density evolves as
(with $a_{\rm now}=1$)
\be
\frac{(\Omega-1)_{\rm now}}{(\Omega-1)}=\frac{H^2a^2}{H^2_{\rm now}}~.
\ee
Let us assume $a(t)\propto t^q$ so that $H\propto q/t$. For (adiabatic) radiation domination, $q=1/2$, while for
matter domination $q=2/3$.
Then we find that at earlier times
\be\label{omegafinet}
(\Omega-1)=\left(\frac{t}{t_{\rm now}}\right)^{2(1-q)}(\Omega-1)_{\rm now}~.
\ee
According to the WMAP observations \cite{wmap}, today $\Omega$ differs from 1 by at most a few per cent. Hence (\ref{omegafinet}) implies a considerable fine-tuning
of the initial value of $\Omega$. For example, let us consider the electroweak phase transition at $t_{EW}\simeq 10^{-11}$ s and
assume for simplicity radiation domination before $t=380\ 000$~yrs followed by matter domination up to $t_{\rm now}=13.7\times 10^9$ yrs.
Then one finds that $\Omega-1$  should have been smaller than today by a factor of $7.5\times 10^{-28}$. The existence of dark energy
does not change the conclusion in any qualitative way.

Such fine-tuning obviously calls for a dynamical explanation, a process that would automatically yield the initial condition
$\Omega=1$. Cosmological inflation solves this problem by assuming that in the very early universe, the energy of the vacuum
$\rho_\Lambda$ was constant and much bigger than any other energy form, including the curvature. A universe with constant vacuum
energy is called the de Sitter universe, where the Friedmann equation (\ref{friedmanni}) would read
\be\label{desit}
H^2=H_0^2-\frac{k}{a^2}~,
\ee
with $3H_0^2=8\pi G\rho_\Lambda$. The solution to (\ref{desit}) at late times behaves as $a(t)=a(t_0)\exp(H_0t)$. The curvature terms
$k/a^2\to 0$ while $\Omega-1\propto\exp(-2H_0t)$. Eventually the vacuum energy should decay, but if the lifetime is long enough,
the initial condition problem is solved as $\Omega$ is driven exponentially close to 1.

Vacuum energy solves also another initial-condition problem, the homogeneity of the
CMB. We see photons arriving at us from every direction with temperatures that are almost exactly equal. They were created
at the (somewhat illogically named) recombination at redshift $z\simeq 1100$ when the ambient plasma became transparent
to photons. The problem is that at the recombination time there were thousands of regions that apparently had not had time
to be in causal contact. This can be seen as follows.
Photons travel along geodesics $ds^2=0$, and taking
the flat ($\Omega=1$) FRW space for simplicity, we thus write (defining the $x$-axis to lie along the photon trajectory)
$ds^2=dt^2-a^2dx^2=0$.
The physical distance a photon travels is then given by
\be\label{photphysd}
d_\gamma(t)=a(t)\int dx=a(t)\int_0^tdta^{-1}=\frac{t^qt^{1-q}}{1-q}~,
\ee
where we have again assumed that $a\propto t^q$. A causal volume at the recombination time $t_{RC}=380\ 000$~yrs is $V\sim d_{RC}^3=(2t_{RC})^3$.
The volume of the visible universe today is roughly (neglecting dark energy)
$V_U\sim (t_{\rm now}/(1-\frac 23))^3$ and at recombination was smaller by a factor of $z^3$ but still much larger than a single
causal volume. Thus one finds that at recombination our universe contained about $10^5$ acausal regions.

Let us observe that in contrast to (\ref{photphysd}), in the de Sitter universe photons travel the physical distance
\be
d_\gamma(t)=a(t)\int dx\simeq \frac{1}{H_0}e^{H_0t}~~,t\to\infty,
\ee
so that two observers separated by an initial distance $d_i$ will at late times find themselves
at a distance $d(t)=d_i e^{H_0t}$. They cannot communicate if $d(t)>d_\gamma(t)$
or if
$d_i>{1}/{H_0}\equiv d_H$. Thus de Sitter space has an event horizon, a radial distance from the observer beyond
which no information can be gathered. The horizon appears because the expansion is superluminal: photons cannot
keep up with the exponential expansion rate.

Assume now that the de Sitter expansion stops at $a_{\rm end}$ and, in the sudden decay approximation, the vacuum energy $\rho_\Lambda$ is
converted instantaneously to radiation with
\be\label{suddend}
\rho_\Lambda=\frac{3H_0^2}{8\pi G}\sim T_0^4~,
\ee
where $T_0$ is the temperature of the universe after the conversion.
This would then mean the beginning of the `normal' Big Bang expansion with $a\sim t^{1/2}$. However,
the preceding superluminal expansion has not been without consequences:
\bi
\item two points $A$ and $B$ initially in causal contact ($d_{AB}(t_0)<H_0^{-1}$) will lose causal contact
because of the exponential expansion when $d_{AB}(t)>H_0^{-1}$ ($A$ is said to have left $B$'s horizon);
\item once `normal' hot Big Bang expansion begins, the horizon
grows as $d_H\sim t$ while $d_{AB}(t)\sim t^q$ with $q<1$ so that
eventually  $d_H>d_{AB}(t)$; hence $A$ and $B$ will again come into causal contact ($A$ is said to re-enter $B$'s horizon).
\ei

Thus, according to inflation, while it seems that there are acausal regions in the CMB sky, they actually have
been in causal contact in the very beginning.
We should require that every point in the CMB sky as seen today had initially
been in causal contact; i.e., the size of the visible universe at the beginning
of inflation is $d_U<H_0^{-1}$. It has then been stretched by expansion of the universe
as
\be
d_U=\frac{a_{\rm now}}{a_{\rm end}} e^{H_0\tau}H_0^{-1}=\frac{T_0}{T_{\rm now}}e^NH_0^{-1}
\ee
where $\tau$ is the duration of inflation and $N=H_0\tau=\ln(a(t_{\rm end})/a_i)$ is the number of e-folds (here we neglect the
details of the expansion history). The actual observed size of the universe today is very roughly
$d_{\rm now}\sim 3/2H_{\rm now}$ (this does not account for dark energy) so that requiring
$d_U>d_{\rm now}$ translates into
\be
N>\ln\left[\frac{d_{\rm now} T_0}{d_UT_{\rm now}}\right]\sim 60
\ee
for reference values $ T_0\sim 10^{15}$ GeV, $T_{\rm now}\simeq 3$~K, $H_{\rm now}\simeq 70$  km~s$^{-1}$/Mpc.
Thus, if inflation lasted more than about 60 e-folds, both the fine-tuning of $\Omega$ and the homogeneity
of CMB find a natural solution. Note that because of the exponential expansion, during inflation the universe becomes
essentially empty.

\section{The inflaton}

\subsection{Equation of state of a scalar field}
If vacuum energy dominates, from the continuity equation (\ref{conteq}) we find the equation of state
\be
p\simeq -\rho~.
\ee
Eventually the vacuum energy should decay and
provide us with the beginning of the hot Big Bang. The simplest way to achieve this is to assume the
existence of a singlet scalar field $\phi$, called the inflaton, with a potential $V(\phi)$ which is of a specific type.
In curved spacetime a scalar field theory is given by the action
\be
S=\int d^4x\sqrt{-g}{\cal L[\phi]}~,
\ee
where $g={\rm det}\ g_{\mu\nu}=-a^6$ for the flat FRW metric (\ref{frwmetric}), while the Lagrangian reads
\be\label{lagr}
{\cal L}=\frac 12g^{\mu\nu}\partial_\mu\phi\partial_\nu\phi-V(\phi)
=\frac 12\dot\phi^2-\frac{1}{a^2}(\nabla\phi)^2-V(\phi)~.
\ee
The energy-momentum tensor for a general Lagrangian reads
\be
T^{\mu\nu}=\frac{\partial{\cal L}}{\partial(\partial_\nu\phi)}\partial^\mu\phi-g^{\mu\nu}{\cal L}
\ee
so that by treating the scalar field as a perfect fluid, we find from (\ref{tmunu}) that
\be
\rho=T_{00}=\dot\phi^2-\left[\frac 12\left(\dot\phi^2-\frac{1}{a^2}(\nabla\phi)^2\right)-V\right]
=\frac 12\dot\phi^2+\frac{1}{a^2}(\nabla\phi)^2+V~.
\ee
We define  comoving pressure $p$ as
\begin{eqnarray}
a^2p&=&T_{kk}=(\partial_k\phi)^2+a^2\left[\frac 12\left(\dot\phi^2-\frac{1}{a^2}(\nabla\phi)^2\right)-V\right]\nn\\
&=&a^2\left(\frac 12\dot\phi^2-V\right)-\frac{1}{6}(\nabla\phi)^2~,
\end{eqnarray}
where the factor $1/6$ comes about because we assume an isotropic universe. Note that
the spatial gradients in $\rho$ and $p$ tend to die away with expansion; this (to some extent) justifies
the assumption of a flat metric in (\ref{lagr}). Thus we find that
\be
\frac{p}{\rho}\simeq\frac{\frac 12\dot\phi^2-V}{\frac 12\dot\phi^2+V}\simeq -1
\ee
if $\dot\phi^2\ll V$. In other words, if the kinetic energy of the scalar field is much
smaller than the potential energy, one can have a de Sitter-like period of exponential
expansion.

The equation of motion for $\phi$ is the Euler-Lagrange equation:
\be\label{eqm1}
0=\frac{\partial\sqrt{-g}{\cal L}}{\partial\phi}-\partial_\mu\frac{\partial\sqrt{-g}{\cal L}}{\partial\partial_\mu\phi}
=\ddot\phi+3H\dot\phi-\frac{1}{a^2}\nabla^2\phi+V'~.
\ee
If we require $\dot\phi^2\ll V$, then we should also require $\dot\phi\ddot\phi \ll \dot V=V'\dot\phi$ or $\ddot\phi \ll V'$.
This is called \emph{slow roll}.

\subsection{Motion of the inflaton}
Let us now assume that the inflaton gradients can be neglected and we may focus on the homogeneous
field $\phi=\phi(t)$. The equation of motion (\ref{eqm1}) reads then
\be
\ddot\phi(t)+3H\dot\phi(t)+\Gamma \dot\phi(t) + V'(\phi(t))=0
\ee
where we have added by hand the decay width $\Gamma$. (In field
theory decay width is the imaginary part of self-energy: $\Gamma=2\ {\rm Im}\ E$.)
Decay can begin only when $\Gamma\gtrsim H(t)$.
Let us assume that near $\phi\simeq 0$ we may expand the potential as
\be
V=V_0-a\phi-b\phi^2-\dots
\ee
where $a$ and $b$ area some parameters. The potential is very flat if $a,~b$ are very small. It then follows
that the Hubble rate is given by $H^2=H_0^2\simeq 8\pi GV_0/3$, where $H_0$ is constant.
Assuming that initially $\phi\simeq 0$, $\dot\phi\simeq 0$ we find that field motion is slow with
$\ddot\phi\ll |V'|$, whence the equation of motion can be written simply as
\be\label{slowrolleqm}
3H_0\dot\phi\simeq -V'~.
\ee

Around the true vacuum $\phi=\phi_*$ we may write the equation of motion as (assuming again that $\Gamma\ll  H(t)$)
\be
\ddot\phi+3H\dot\phi+\frac 12m_\phi^2(\phi-\phi_*)^2\simeq 0~,
\ee
where $m_\phi$ can be called the physical mass of the inflaton, which typically
is much bigger than $V''$ during inflation.
Writing $\xi=A(t)(\phi-\phi_*)$ one finds for the amplitude $A(t)\propto a^{-3/2}$, while
averaging over one oscillation cycle the mean pressure is found\cite{turner} to be $\langle p\rangle =0$.
Hence a harmonically oscillating field behaves as cold matter with $\rho\propto a^{-3}$ and
$a\propto t^{2/3~}$.

The inflaton oscillations will continue until $H\simeq\Gamma$, whence the inflaton
starts to decay into some (relativistic) particles. The decay products will eventually thermalize
among themselves. The energy density stored in the oscillations
is transformed at $t=t_{\rm dec}$ into radiation so that [see also (\ref{suddend})]
\be
\rho_{\rm thermal}=\frac{g_*\pi^2}{30}T_{\rm RH}^4=\rho_{\rm end}\left(\frac{a_0}{a(t_{\rm dec})}\right)^{3}~,
\ee
where $\rho_{\rm end}\simeq V_0$ is the inflaton energy density at the end of inflation and
$T_{\rm RH}$ is the \emph{reheat temperature}. The decay time is given by the condition $t_{\rm dec}=\Gamma^{-1}=H(t_{\rm dec})$.

Thus we arrive at the following scenario:
\bi
\item initially the inflaton is in the flat part of the potential, slowly rolling; the universe
is expanding (quasi)exponentially;
\item as the field gathers speed, slow-roll conditions no longer hold and inflation ends;
\item inflaton begins to oscillate about the true minimum $\phi_*$; the universe is
expanding as if dominated by cold matter;
\item when $H\simeq\Gamma$, the inflaton starts to decay and (re)heats the universe, which then starts expanding in a radiation-dominated phase
\ei

The decay of the inflaton field can in principle proceed also in a non-perturbative manner through so-called parametric
resonance \cite{parares}. By coupling the inflaton to other scalar fields, one may arrange for a situation where the
decay products are generated in bursts while the inflaton oscillates past the minimum of the potential. This results
in a much more efficient reheating than the conventional perturbative decay.

\subsection{Slow-roll parameters and the number of e-folds}
\vskip10pt
It is convenient to define slow-roll parameters that characterize the inflaton
potential during inflation. The spectral index of the perturbations can also be
expressed in terms of the slow-roll parameters, as we will see.
Since $V\gg \dot\phi^2$, from the slow-roll equation of motion (\ref{slowrolleqm})
we deduce that
\be
1\gg \frac{1}{V}\left(\frac{V'}{3H}\right)^2\simeq \frac{1}{V}\frac{(V')^2}{8\pi G V_0}~.
\ee
Let us therefore define the slow-roll parameter $\epsilon$ as
\be\label{epsdef}
\epsilon \equiv\frac{M_P^2}{16\pi}\left(\frac{V'}{V}\right)^2\equiv\frac{M^2}{2}\left(\frac{V'}{V}\right)^2\ll 1
\ee
where $M=M_P/\sqrt{8\pi}$ defines the reduced Planck mass.
Taking the derivative $d(V')^2/d\phi$ one finds that $V''/V\ll 16\pi/M_P^2$ so that one may define
\be
\eta\equiv\frac{M_P^2}{8\pi}\frac{V''}{V }\equiv M^2\frac{V''}{V }\ll 1~.
\ee
The slow-roll conditions become violated and inflation ends when $\epsilon, \eta\simeq 1$.

The number of e-folds is then given by
\be\label{efolds1}
N=\ln\frac{a(t_{\rm end})}{a(_i)}= \int_{t_i}^{t_{\rm end}}dt H(t)~,
\ee
where $t_{\rm end}$ corresponds to $\epsilon(\phi),~\eta(\phi)\simeq 1$. Using the slow-roll equation
(\ref{slowrolleqm}) we may write (\ref{efolds1}) as
\be
N=-\frac{1}{M^2}\int_{\phi_i}^{\phi_{\rm end}}d\phi\frac{V}{V'}~.
\ee
Comparing this with (\ref{epsdef}) we see that the required large number of e-folds can be obtained if the slow-roll parameter $\epsilon$ is small enough.
\section{Inflationary perturbations}
\subsection{Evolution of field perturbations}
Like any quantum field, the inflaton is subject to fluctuations. Hence, we should write
the inflaton as
\be\label{fullphi}
\phi(x,t)=\phi_0(t)+\delta\phi(x,t)~,
\ee
where $\phi_0(t)$ is the homogeneous part which is treated here as the background field that is the solution
to the slow-roll equation (\ref{slowrolleqm}), and $\delta\phi(x,t)$ is the (small) perturbation.
Since during
inflation the energy density $\rho \sim V$, field fluctuations source
also perturbations of energy density with $\delta\rho\propto \delta V(\phi(x,t))=V'(\phi_0)\delta\phi(x,t)$.
Eventually, such perturbations can be observed in the microwave sky as temperature fluctuations. Given the
inflationary model, we are thus in a position to calculate the spectrum of the CMB fluctuations.
This requires two things: 1) assumptions about the initial conditions for the field perturbation, and
2) understanding the evolution of the perturbation.

Assuming that the perturbation $\delta\phi(x,t)$ is small, after substituting (\ref{fullphi})
to the equation of motion (\ref{eqm1}) we find that to lowest order $\delta\phi(x,t)$ obeys
\be\label{perteqm}
\delta\ddot\phi_k+3H(\phi_0)\delta\dot\phi_k+\left(\frac{k^2}{a^2}+m_\phi^2\right)\delta\phi_k=0~,
\ee
where we have moved to Fourier space.

Obviously, for massless fields and for long-wavelength
fluctuations with $k\to 0$, one finds that $\delta\phi_k\to$ const.
This means that once beyond the horizon, the field perturbation freezes.
Because it has lost causal contact with the horizon patch where it originated,
it can no longer be modified by local physics. This is of course good news as
it implies that the spectrum of perturbations can be computed if we only knew
its amplitude at the time it crosses the horizon.

For a perturbation well within the horizon, $k/a\gg H$. Moreover,
the slow-roll condition requires that $m_\phi\ll H$.
Hence at early times we may write (\ref{perteqm}) as
\be\label{perteqm2}
\delta\ddot\phi_k+3H(\phi_0)\delta\dot\phi_k+\frac{k^2}{a^2}\delta\phi_k=0~.
\ee
Hence we have an evolution equation for the perturbation, but what is the initial condition for the Fourier mode $\delta\phi_k$?
The sensible assumption is that at very small distance scales, well inside the horizon, curvature can be neglected and that
locally space looks Minkowski. It is then sufficient to consider
quantum fluctuations in empty flat space\footnote{The concept of vacuum in curved space is not unproblematic since
the particle content for observers in different frames can be different. This particular choice is called the Bunch--Davies vacuum.}
and quantize the free inflaton field in the usual manner with
$\phi_k=w_k(t)a_k+w_k^*(t)a_{-k}^\dagger$ where the amplitude is given by
\be\label{wk}
w_k=V^{-1/2}\sqrt{\frac{1}{2E_k}}\exp(-iE_kt)~,
\ee
while
$[a_k,a_{k'}^\dagger]=\delta_{kk'},~[a_k,a_{k'}]=0$. The vacuum is
then defined through the usual condition
$a_k|0\rangle=0$ with $a_k^\dagger|0\rangle \propto|k\rangle$.

It is easy to see that in the vacuum the inflaton VEV is zero whereas the variance is given by
\be\label{qvariance}
\langle 0|\phi^2|0\rangle=\sum_k|w_k|^2~.
\ee
This suggests that
the initial perturbation $\delta\phi_k$ at $t\to-\infty$ should be identified with the
root-mean-square of the the variance (\ref{qvariance}).
Thus, we should find the solution of (\ref{perteqm2}) with $\delta\phi_k=w_k$ and
$w_k(t\to -\infty)$ given by (\ref{wk}) in a box of size $L$. Here we will
ignore the variation of $H$ and assume simply that $H(\phi_0(t))\simeq H_0$; this is consistent
with the slow-roll assumption. It is straightforward to show that
the solution is given by
\be
w_k=\frac{1}{L^{3/2}}\frac{H}{(2k^3)^{1/2}}\left(i+\frac{k}{aH}\right)\exp{\left(\frac{ik}{aH}\right)}
\ee
which reduces to the flat space result\footnote{This can be seen by expanding $k/(aH)$ about some reference time $t_0$
by writing ${k}/({aH})={k}/({aH})|_{t=t_0}+{d}[{k}/({aH})]/dt|_{t=t_0}(t-t_0)$.} (\ref{wk}) in the limit $t\to-\infty$.
The initial field perturbation (\ref{qvariance}) is Gaussian, but later evolution can, depending on the model, also generate small non-Gaussian
features.

At horizon exit $t=t_*$, we have $k=aH$ so that the perturbation amplitude is then given by
\be\label{pertamplt}
|w_k|^2=\frac{H^2(t_*)}{2L^3k^3}~.
\ee
The power spectrum $P(k)$ associated with a fluctuation $\delta_k$ is
defined as
\be
P(k)=\frac{V}{(2\pi)^3}k^2d\Omega\langle|\delta_k|^2\rangle=\frac{Vk^3}{4\pi^2E_k}\langle|\delta_k|^2\rangle~,
\ee
where $V$ is the volume of the box.
Thus at the horizon exit we find the power spectrum of the inflaton fluctuations as
\be\label{powerspec}
P(k,t_*)=\frac{Vk^3}{4\pi^2E_k}\frac{H^2(t_*)}{2L^3k^3}=\left(\frac{H}{2\pi}\right)^2_{k=aH}~.
\ee
Note that the box-size dependence cancels. Equation (\ref{powerspec}) is a central
result for cosmological inflation.

Because we are assuming slow roll, the value of $H$ is almost constant in time. It then follows that the
amplitude of the perturbation (\ref{pertamplt}) is almost the same for any arbitrary scale $k$ at the
horizon exit so that the observed spectrum should be (almost) scale independent.

\subsection{From field perturbations to temperature fluctuations}
Inflaton perturbations generate density perturbations, or more generally,
perturbation in the energy-momentum tensor, which is the source for
the metric. Hence, there will also be metric perturbations.
In the so-called Newtonian gauge the perturbed metric reads, neglecting
now vector and tensor perturbations (gravitational waves),
\be
ds^2=(1+2\Phi)dt^2-a^2(t)(1+2\Psi)dx^2~,
\ee
where $\Psi$ and $\Phi$ are called the Bardeen potentials; for a perfect non-viscous fluid $\Psi=\Phi$.

However, in general relativity one
is free to change coordinates in any way one wishes by performing a general coordinate
transformation $g_{\mu\nu}\to g'_{\mu\nu}$. Therefore,
one could `gauge away' the density perturbation by making a coordinate transformation such that $\delta\rho(x,t)=0$ everywhere (this is called `constant
density hypersurface'). Obviously, we need some gauge-invariant description of the perturbation in order
to really be able to decide what is observable.

Without going into details, let us state that one can show\cite{gaugeinv} that the comoving curvature perturbation
\be
{\cal R}=-\frac{H}{\dot\phi}\delta\phi
\ee
is
both gauge invariant and also remains constant outside horizon. It is related to the
Bardeen potential by $\Psi=\frac 35{\cal R}$. In the literature one often uses also
the curvature perturbation $\zeta$, which outside the horizon is simply given by
$\zeta=-{\cal R}$.

CMB observations yield correlators such as
\be
\langle \frac{\delta T}{T}\frac{\delta T}{T}\rangle_k\equiv P_T(k)~,
\ee
or the CMB temperature power spectrum.
One can show that for large angular distances
${\delta T}/{T}=-\frac 13\Psi$, which is called the
Sachs--Wolfe effect. For small angular distances (with multipoles $l\gtrsim 50$) one needs also to consider
hydrodynamical effects in the photon--baryon plasma, which give rise to the well-known acoustic oscillations in
the CMB power spectrum.

Here, let us focus on the Sachs--Wolfe effect alone. We need  to
compute the power spectrum for the comoving curvature perturbation:
\be
P_{\cal R}(k)=\langle {\cal R}_k{\cal R}_k
\rangle=\frac{H^2}{\dot\phi^2}\langle \delta\phi_k\delta\phi_k
\rangle =\frac{H^2}{\dot\phi^2}\left(\frac{H}{2\pi}\right)^2_{k=aH}.
\ee
Using the slow-roll condition (\ref{slowrolleqm}) and the
definition of the slow-roll parameter $\epsilon$ (\ref{epsdef}) one
then finds
\be P_{\cal
R}(k)=\frac{1}{24\pi^2}\frac{V}{M^4}\frac{1}{\epsilon}~.
\ee
The
amplitude of the perturbation, often called the COBE normalization, is conventionally defined as
as $\delta_H^2=4P_ {\cal R}/25$ with the observed value $\delta_H=1.9\times
10^{-5}$. This constrains the scale of the inflaton potential to be $(V/\epsilon)^{1/4} =0.027M$.
Thus, for a very low scale model, say, with $V\simeq M_W^4$, one would require an extremely flat potential with $\epsilon$ very
close to zero.

The CMB spectrum is specified by its amplitude and by the spectral index $n$.
Purely phenomenologically we may thus write
\be
P_{\cal R}=Ak^{n-1}~,
\ee
where $n-1={d\ln P_{\cal R}}/{d\ln k}$. Thus, if the spectrum were exactly scale independent, we would find $n=1$.
Because of the slow roll, we expect the deviation from scale independence to be small.
Let us now compute the spectral index. At the horizon exit $k=aH$ so that the differential
$d\ln k={da}/{a}+{dH}/{H}\simeq Hdt$ since during inflation $H\simeq$ const.
From the slow-roll condition (\ref{slowrolleqm}) we find that
$dt=-3Hd\phi/V'$ so that
\be
\frac{d}{d\ln k}=-\frac{V'}{3H^2}\frac{d}{d\phi}=-M^2\frac{V'}{V}\frac{d}{d\phi}~.
\ee
Then ${d\epsilon}/{d\ln k}=4\epsilon^2-2\epsilon\eta$ and
\be
n-1=2\eta-6\epsilon~.
\ee
Hence given the inflaton model, we are able to predict both the amplitude and the spectral
index of the CMB temperature fluctuation.
WMAP7\cite{wmap} yields $1-n=0.037 \pm 0.014$ (for CMB alone). Hence typically (but not necessarily) $\epsilon\sim {\cal O}(0.01)$. Other
typical  values would then be the scale of the potential during inflation
$V^{1/4}\simeq 10^{16}$ GeV and the Hubble rate during inflation
$H\simeq 10^{14}$ GeV. However, all these number are very much model dependent.

Scalar-field-driven inflation is a generic idea but as of yet, there is no compelling, particle physics
motivated theory of inflation. Instead, there exists a vast number of different models. There are models
with many inflaton fields, models based on extra dimensions, models based on the Higgs field with a non-minimal
coupling to gravity, models where the the superluminal expansion and the primordial perturbation are generated
by different fields (for reviews of the various inflaton models, see Ref.~\cite{reviews}). The present observations of the
CMB spectral index yield some interesting constraints on the models, but perhaps the best hope for a
decisive test of the various models could be provided by the ESA Planck Surveyor Mission
\cite{Planckportal}, which is expected to put a stringent limit to the non-Gaussianity of the primordial
perturbation, or perhaps even observe it. Should that happen, many models would immediately be ruled out.


\begin{thebibliography}{99}
\bibitem{wmap} WMAP results are available at the URL lambda.gsfc.nasa.gov.
\end{thebibliography}
\end{document}